\documentclass[fleqn]{annalen}

\usepackage{graphics}

\begin{document}


\newcommand{\volume}{?}              
\newcommand{\xyear}{2002}            
\newcommand{\issue}{December}               
\newcommand{\recdate}{6 November 2002}    
\newcommand{\revdate}{dd.mm.yyyy}    
\newcommand{\revnum}{0}              
\newcommand{\accdate}{12 November 2002}    
\newcommand{\coeditor}{B.~Kramer}           
\newcommand{\firstpage}{1}           
\newcommand{\lastpage}{?}            
\setcounter{page}{\firstpage}        


\newcommand{\ie}{\mbox{i.\ e.\ }}
\newcommand{\eg}{\mbox{e.\ g.\ }}
\newcommand{\nb}{\mbox{N.\ B.\ }}
\newcommand{\cf}{\mbox{c.\ f.\ }}
\newcommand{\etc}{\mbox{etc.}}
\newcommand{\be}{\begin{eqnarray}}
\newcommand{\en}{\end{eqnarray}}
\newcommand{\no}{\nonumber}
\newcommand{\new}{\no\\ && \qquad}
\newcommand{\hc}{\mbox{\it h.\ c.\ }}
\newcommand{\mc}{\mathcal}


\newcommand{\keywords}{pyrochlore lattice,fractional charge,heavy fermions} 
\newcommand{\PACS}{74.40.Cx,71.27+a,71.10-w}


\newcommand{\shorttitle}{P.\ Fulde et al., Fractional charges in pyrochlore
lattices} 

\title{Fractional charges in pyrochlore lattices}

\author{Peter Fulde$^{1}$, Karlo Penc$^{1,2}$, and Nic Shannon$^{1}$} 

\newcommand{\address}
  {$^{1}$Max-Planck-Institut f\"ur Physik komplexer Systeme,
   \\ \hspace*{0.5mm} N\"othnitzer Stra\ss e 38
   \\ \hspace*{0.5mm} 01187 Dresden, Germany \\ 
   $^{2}$Research Institute for Solid State Physics and Optics
   \\ \hspace*{0.5mm} H--1525 Budapest, P.O.B. 49, Hungary}

\newcommand{\email}{\tt fulde@mpipks-dresden.mpg.de} 
\maketitle

\begin{abstract}
A pyrochlore lattice is considered where the average electron number of
electrons per site is half--integer, concentrating on the
case of exactly half an electron per site.  
Strong on-site repulsions are assumed, so that all sites are 
either empty or singly occupied.  When there are in addition 
strong nearest--neighbour repulsions, a tetrahedron rule comes
into effect, as previously suggested for magnetite.  
We show that in this case, there exist excitations with 
fractional charge $\pm e/2$.  
These are intimately connected with 
the high degeneracy of the ground state in the absence of kinetic 
energy terms.  
When an additional electron is inserted into the system, 
it decays into two point like excitations with 
charge $-e/2$, connected by a Heisenberg spin chain 
which carries the electron's spin.  
\end{abstract}

\section{Introduction}

The experimental observation of heavy-fermion behaviour of LiV$_2$O$_4$
\cite{KJSBMMGGMG,KoJoMi,Johnston} has
drawn attention to pyrochlore lattice systems with half--integer valency of the
involved ions. 
For example, the average valency of V in LiV$_2$O$_4$ is $+3.5$,
and the same is true for Ti in LiTi$_2$O$_4$. Therefore, the average 3d
electron number is d$^{1.5}$ in the first case, and d$^{0.5}$ in 
the second one.  
From LDA band--structure calculations 
\cite{EyHoLoRi,MaFuMa,AnKoZoPrJuRi,SiBlScMa} it is known that the 
conduction bands have 3d-t$_{2g}$ character, and are well separated 
from higher energy valence electron states.
However the LDA effective electron mass found for LiV$_2$O$_4$
is a factor of twenty--five smaller than the quasiparticle mass 
inferred from specific heat and spin susceptibility data. 
This is a sign of strong electronic correlations. 
On--site Hubbard $U$ interactions alone 
are not sufficient to explain the large measured quasiparticle mass. 
They merely reduce the atomic configurations of the V ions to 
3d$^1$ and 3d$^2$, \ie, they exclude 3d$^0$, 3d$^3$ configurations, \etc 
In order to obtain a sufficiently high density of low--energy 
excitations one must therefore include correlations between
neighbouring sites.  The nearest--neighbour interactions are minimised if,
for each of the corner sharing tetrahedra which make up the pyrochlore
lattice, there are two 3d$^1$ and two 3d$^2$ configurations. 
This so--called ``tetrahedron rule'', which is implicit in Verwey's
treatment of the metal--insulator transition in the spinel
Fe$_3$O$_4$ \cite{VerHaa}, was first stated explicitly by 
Anderson \cite{Anderson}. 
The number of configurations which obey the tetrahedron rule grows 
exponentially with the number of tetrahedra, so in the absence of
any perturbation (such as the electrons' kinetic energy) which 
selects between these states, the ground state of pyrochlore systems with
nearest--neighbour interactions and half--integer valency is extremely  
degenerate.  An important property of these degenerate ground--state 
configurations is that each of them consists of chains of 3d$^1$ 
and 3d$^2$ sites in the case of LiV$_2$O$_4$ \cite{FuYaZvGr}, 
and of empty 3d$^0$ and 3d$^1$ sites in the case of
LiTi$_2$O$_4$. 
Spin excitations in those chains can give raise to a linear low
temperature specific heat C(T)=$\gamma$T with a large $\gamma$ 
coefficient\cite{FuYaZvGr}. 
Because of the special features of one--dimensional systems, these
spin excitations can be described by either bosons or fermions
\cite{Schulz}.  

In this paper we set to explore the effects of a small kinetic energy
term on a system obeying the tetrahedron rule.
For simplicity we shall limit our discussion to the case 
of an average 3d$^{0.5}$ count, \ie, to states with one electron 
for every two lattice sites.  
We note that the calculations 
are not intended to apply to LiTi$_2$O$_4$, for which no
heavy--fermion behaviour has been observed, but should instead 
be considered as having model character.  
They are intended to demonstrate special 
features of excited states in charge--frustrated lattices.  
In order to be better able to vizualize states obeying the 
tetrahedron rule, we further simplify the problem by 
considering a two--dimensional checkerboard lattice instead of 
the original three--dimensional pyrochlore lattice. 
This is essentially a special projection of the pyrochlore
lattice onto a plane, and all of the results we discuss
apply equally to three--dimensional pyrochlore systems.

\section{Model Hamiltonian}

For a pyrochlore lattice with t$_{2g}$ electrons, the full model
Hamiltonian is of the form  \cite{KleMed}
\be
{\mc H} &=& \sum^{12} _{\nu=1 \atop {\bf k}, \sigma} \epsilon_\nu({\bf k}) 
   a^+_{{\bf k} \nu \sigma} a_{{\bf k} \nu \sigma} 
   + (U-2J) \sum_{ia} n_{a \uparrow} (i)
   n_{a \downarrow} (i)\nonumber\\
   &+& \frac{1}{2} \left(U-\frac{J}{2}\right) \sum_{a \neq a' \atop i} 
   n_a(i) n_{a'}(i) - J \sum_{a \neq a' \atop i} 
   {\bf s}_a (i) {\bf s}_{a'}(i)\nonumber\\ 
   &+& \bar{J} \sum_{\langle ij \rangle} {\bf S} (i) {\bf S} (j) 
   + V\sum_{\langle ij \rangle} n (i) n (j) ~~~.
\label{eq:HsumAT}
\en
The first term describes the kinetic energy of the t$_{2g}$ electrons
which form twelve bands with index $\nu$ because of the four sites per unit
cell. 
The following three terms are due to Coulomb repulsions of d electrons 
on a transition--metal ion site $i$ occupying orbitals $a$ and $a^{\prime}$. 
Here $U$ is the direct and $J$ the exchange part of the on-site Coulomb 
interaction; the latter enforces Hund's first rule when several 
d electrons occupy the same site. 
We shall assume that U is large enough to ensure that each ion 
fluctuates between two valencies only.
The remaining two terms describe the exchange $\bar{J}$ and 
direct Coulomb $V$ interactions between neighbouring sites.
These interactions are {\it frustrated} by virtue of the 
geometry of the pyrochlore lattice, and this is an essential 
prerequisite of our theory.  In particular, it is the nearest neighbour 
repulsion $V > 0$ which singles out states obeying the tetrahedron rule.

In order to first discuss excitations associated with charge degrees 
of freedom, we begin by considering the reduced Hamiltonian
\be
{\mc H}_0 &=& -t \sum_{\langle ij \rangle} \left\{ 
   f^{\dagger}(i) f(j) + \hc \right\}
   + V \sum_{\langle ij \rangle} n^{f}(i) n^{f}(j)
\label{eq:H0tsum}
\en 
where $n^f(i) =  f^{\dagger}(i) f(i)$.
This refers to a system with an average d-electron count of $0.5$ 
and one orbital per site.  Because of the assumed large value
of $U$, only empty (d$^0$) and singly occupied (d$^1$) sites are
considered.  Spinless fermions are assumed, which corresponds to a 
full spin polarisation of the electrons. 
Charge conservation requires the subsidiary condition 
\be
\label{eq:sumTau}
 \sum_i n^{f}(i) = N/2
\en 
for a system with N lattice sites, while in the limit 
$V \to \infty$ the tetrahedron rule requires that 
\be
\sum_{\nu =1}^4 \left( n^{f}(i_{\nu}) - \frac{1}{2} \right) &=& 0
\label{eq:sum4v}
\en
where $\nu$ denotes the four different sites within {\it any} 
given tetrahedron $i$.
In one dimension, this Hamiltonian can be mapped onto
an anisotropic Heisenberg model, and can be solved 
exactly \cite{ovchinnikov}. 
In this case one finds a gap-less metallic excitation spectrum 
for $V < 2t$, and a gapped insulating state for larger values of $V$. 

For a pyrochlore lattice no such exact results are available. 
However, we expect that the parameter range for the two phases 
remains qualitatively unchanged, since the increased number of nearest 
neighbours enters the hopping and interaction terms in the same way. 

\section{Fractional charge}

\begin{figure}
\centerline{\resizebox{12cm}{5cm}{
\includegraphics{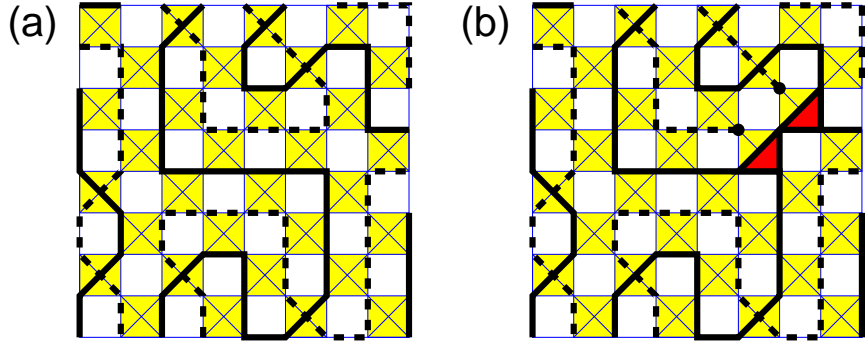}
}}
\caption{\label{fig1}Checkerboard lattice, the thin lines indicate hopping and
interactions connected sites. Thick lines connect sites in a d$^1$
configuration while dotted lines connect empty (d$^0$) sites.
(a) example of a fraction of the lattice in which the tetrahedron rule is
obeyed.  (b) the sample as (a) but with an electron added ($d^0 \to d^1$).
Dots indicate the end of the chain.}
\end{figure}

We consider a two--dimensional checkerboard lattice consisting 
of N sites with a configuration $\mid \Phi_I \rangle$ representing 
one of the degenerate ground states, \ie, a
configuration in which the tetrahedron rule is satisfied (for an example see
Fig. \ref{fig1}a).   It is known that, for the 2D checkerboard lattice, 
the number of these configurations scales as $(4/3)^{3N/4}$ \cite{lieb}.
The wavefunction of $\mid \Phi_I \rangle$ is of the form 
\be
\Phi_I \left(1, ..... , \frac{N}{2}\right) = {\mc A} \prod_{\nu} \phi
\left(\nu(1), ..... , \nu(N_{\nu})\right)
\label{eq:Phi1N2}
\en
where $\sum\limits_\nu N_\nu=N/2$, and the operator ${\mc A}$ 
antisymmetrizes the wavefunction. 
Here $\phi (1, ... , M)$ is the exact ground--state
wavefunction of a Heisenberg chain with an even number M of sites.
The chains have to be arranged in such a way that the tetrahedron 
rule is satisfied by every cross--linked square of the lattice.  

When a charge $-e$ is added to an empty site $j$ we are left with 
two corner--sharing squares with three d$^1$ sites each. 
This is illustrated in Fig. \ref{fig1}b. 
The corresponding state is 
$\mid j \rangle = f^{\dagger}(j)\mid \Phi_I \rangle$. 
The kinetic energy term now permits the five electrons 
within these two squares to move.  
However, the motion has to occur in such a 
way that the number of squares with three d$^1$ sites is conserved.  
The energy barrier to generating states with additional violation 
of the tetrahedron rule is $V$, which has already 
been assumed to be sufficiently large to prevent this from happening.  
And, critically, there is
no way to ``undo'' the violation of the tetrahedron rule
geometrically, simply by moving electrons around the lattice.  
The number of tetrahedra (squares) in which the
tetrahedron rule is violated is therefore a topological invariant 
of the states which we consider.

\begin{figure}
\centerline{\resizebox{12cm}{5cm}{
\includegraphics{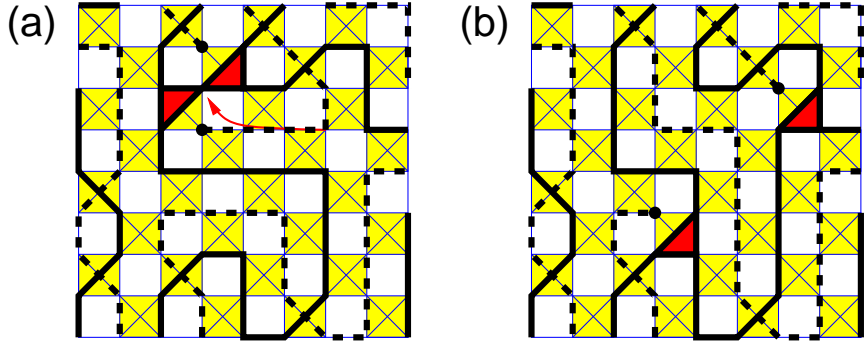}
}}
\caption{
\label{fig2}
(a) The added electron has moved in four steps along a chain of empty
sites. It remains an entity.  (b) An electron from lower triangle in
Fig. \ref{fig1}b has moved along the diagonal. As a consequence 
the excitation has decayed into two with a fractional charge of $-e/2$ 
each.}
\end{figure}

If the added electron moves along the chain of empty sites in
which it was inserted, it retains its integrity and  
the state $\mid j \rangle$ transforms successively into states 
$\mid n \rangle = f^{\dagger}(n)f(j) \mid j \rangle$. 
However, if we instead allow one of the four neighbouring electrons
belonging to chains of occupied sites to move, the 
electron immediately breaks into two disjoint pieces.
These carry fractional 
electric charge $-e/2$ (note that every electron is shared 
by two squares and so contributes a charge $-e/2$ to each of them).  
These two cases are indicated in Figs. \ref{fig2}a,b.

Energy and momentum (as well as topological charge) must be 
conserved by these decay processes.  Therefore if we were to
associate a momentum ${\bf k}$ and energy $E({\bf k})$ with 
the electron which we inserted, this must now be shared between
the fractionally charged particles into which it has decayed
\be
E({\bf k}) = 4V + \epsilon ({\bf k}_1) + \epsilon ({\bf k}_2)
\label{eq:Ekeps}
\en
where $\epsilon ({\bf k})$ is the dispersion of the fractional
charge and ${\bf k}_1 + {\bf k}_2 = {\bf k}$. 
The situation resembles that of spin excitations in a Heisenberg chain. 
There, a spin $1$ spin--flip excitation can decay into two spin $1/2$ 
domain walls (spinons), with the result that for each value of $k$ there
is a continuum of spin excitations.  Clearly, a metal which has a 
charge excitation spectrum with low energy contributions of the 
form of Eq. (\ref{eq:Ekeps}) {\it cannot} be a conventional Fermi liquid.  

We can obtain some feeling for the form of $\epsilon ({\bf k})$ 
by modelling the fractional charge as a free particle on a
checkerboard lattice with a reduced bandwidth due to the hopping
restrictions imposed by the tetrahedron rule.  
In the case of hopping between nearest--neighbour 
sites only, this would be a band with 
dispersion 
$\epsilon ({\bf k}) = -t [\cos(k_x a) + \cos(k_y a) ]$, 
where $a$ is the lattice constant.  Without the hopping restrictions the
prefactor would be twice as large, and the band two--fold degenerate; 
if the diagonal hopping elements were also restored one of the two bands 
would become dispersionless.
The corresponding free electron dispersion for the pyrochlore 
lattice, including flat bands, is given in \cite{IsoMor}.

Physically, the fractionally charged excitations in our model result from 
a back--flow of charge.  When a fractional charge travels along a 
given path, it leaves behind it a ``wake'' of squares 
(tetrahedra in the case of a pyrochlore lattice) which
still obey the tetrahedron rule, but whose 
charge configuration has been modified.
Motion constrained by the tetrahedron rules occurs such
that each ``hopping'' event of an electron is accompanied by a net 
back--flow of a charge $-e/2$. 
A fractional charge of $-e/2$ is therefore carried by the 
moving particles. 
We want to stress that all of these arguments about
fragmentation of charge depend only on the geometry
of the lattice (more precisely, on the geometry of the repulsions $V$), 
and not on its dimensionality or details of the band structure.
Everything which has been demonstrated here for the checkerboard 
lattice therefore also applies to the three--dimensional pyrochlore 
lattice. 

\section{Spin of fractional charges}

When an electron is added to an empty site, a spin 1/2 is added to the
system.  In order to gain some insight into how this spin is
distributed when the electron decays into two fractional 
charges, we now restore a spin degree of freedom to each d$^1$ site
and consider the model
\be
\label{eq:HH0J16}
{\mc H} &=& {\mc P} \left[ -t\sum_{\langle ij \rangle\alpha} 
   \left\{ d^{\dagger}_{\alpha} (i) d_{\alpha} (j) + \hc \right\}
    \right] {\mc P}\no\\
   &&\qquad + V \sum_{\langle ij \rangle} n^d(i) n^d(j)
            + \bar{J} \sum_{\langle ij \rangle} 
              {\bf S} (i) . {\bf S} (j) 
\en 
where $\alpha=\{\uparrow,\downarrow\}$ is the electron 
spin index,
\be 
n^d(i) = \sum_{\alpha} d^{\dagger}_{\alpha} (i) d_{\alpha}(i)
   \quad, \quad {\bf S} (i) = \frac{1}{2}\sum_{\alpha\beta} 
        d^{\dagger}_{\alpha}(i) \vec{\sigma}_{\alpha\beta} d_{\beta}(i) 
   \quad , \no
\en
and 
\be
{\mc P} = \prod_i \left[1 - n^d_\uparrow(i)n^d_\downarrow(i)\right]
\en
is a (Gutzwiller) projection operator ensuring
that no site is doubly occupied.
At exact quarter filling, all configurations which obey the
tetrahedron rule are composed of randomly distributed Heisenberg 
spin chains, which, for periodic boundary conditions, have 
even length.  Only exchange interactions then come into play,
and these select a singlet ground state.

When an electron is added to one of these states
it either connects two adjacent spin chains, as illustrated in 
Fig. \ref{fig1}b, or two neighbouring 
sections of a single closed spin--chain loop.
Let us suppose for a moment that $t=0$, and the electron is 
fixed at this site.
From density--matrix renormalisation group (DMRG) calculations 
it is known \cite{ZhIgFu} that a spin
$1/2$ coupled symmetrically to a single Heisenberg chain 
has a ground state which is a Kramers doublet, and a 
Curie-like susceptibility. 
We expect the same to be true for a
spin $1/2$ coupled symmetrically to two Heisenberg chains as in
Fig. \ref{fig1}b.  Again the frustration of the lattice plays
a key role in determining the nature the states, but this
time it is the spin interactions which are frustrated.
   
\begin{figure}
\centerline{\resizebox{5cm}{5cm}{
\includegraphics{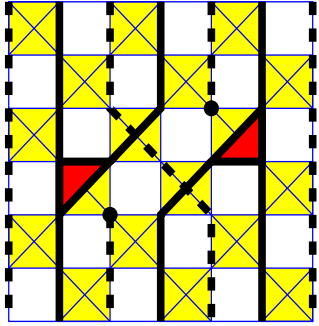}
}}
\caption{A configuration with periodic boundary conditions which demonstrates
that two Heisenberg chains are connected by a third one consisting of an odd
number of sites.
\label{fig3}
}
\end{figure}

If we now restore the kinetic energy term, we once again
have two possible cases for electronic motion.  When
the electron moves along a chain of empty sites, its spin 
interactions are at every step frustrated, and so it can
carry its spin $1/2$ intact as it moves.  But as soon
as the electron decays into two fractional charges, its
spin is no longer localised at one point.
In most cases, the insertion of an electron will 
connect two different spin chains.  The decay 
of the electron converts a single spin 1/2 site into a spin chain 
with an odd number of spin $1/2$ sites coupled symmetrically 
to two Heisenberg chains with an even number of sites.
Since the total number of sites involved is odd, the ground state 
remains a Kramers doublet, as required.  But it is no longer possible
to uniquely assign this spin $1/2$ degree of freedom to a single site.
This situation is illustrated in
Fig. \ref{fig3}.  For better visualisation we have chosen here a special
configuration in which in the absence of the added electron all chains 
have a regular charge order.  This is necessary only to be able to 
illustrate the effect using a small segment of the checkerboard lattice;
global charge order is {\it not} required.

\section{Quantum fluctuations}

In the arguments so far presented we have considered the limit 
$t/V \to 0$ for which repulsive interaction between
electrons on neighbouring sites can be replaced with a simple
``tetrahedron rule''.  If we relax this constraint to
the extent of considering finite (but very large) $V$, quantum
fluctuations about the many degenerate states obeying the 
tetrahedron rule can themselves single out a ground state.  
The problem then falls into the general class of order from disorder
effects, widely studied in the context of frustrated magnetic 
insulators, but largely unexplored for metals.

\begin{figure}
\centerline{\resizebox{5cm}{5cm}{
\includegraphics{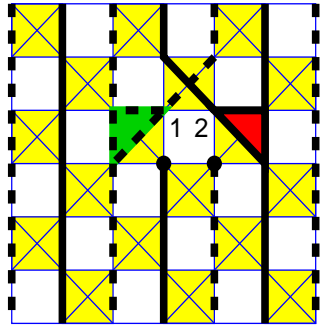}
}}
\caption{Vacuum fluctuation due to an electron hopping from 1 to 2. Two
charges $\pm$e are generated which can propagate freely.
Note that the Heisenberg chains involve even 
numbers of sites.
\label{fig4}}
\end{figure}

Here we briefly discuss two aspects of these quantum fluctuations;
the fact that they also contain fractional charges, and their relevance
for the metal--insulator transition at (smaller) finite $V$.
We start from one of the degenerate ground states $\mid\Phi_I \rangle$ 
which obey the tetrahedron rule, and consider to lowest order the 
effect of the kinetic energy.
At ${\mc O}(t/V)$ this mixes into the wavefunction virtual states 
in which two squares (tetrahedra) violate the tetrahedron rule 
at the net cost of an interaction energy of $V$. 
However overall charge must be conserved, so in this case 
one square contains three d$^1$ sites, while the other 
has three d$^0$ sites.
This is illustrated for the checkerboard
lattice in Fig. \ref{fig4}. 
If this virtual excitation lives long enough for the 
contributing electrons to hop to neighbouring sites it, just 
like an added electron, will decay into two fractional charges.
In this case the two pieces of the excitation 
must carry the opposite sign of electrical charge.
We have already assigned a charge $-e/2$ to the square 
containing three d$^1$ sites; the square 
containing three d$^0$ sites carries a net charge $e/2$,
and the number of squares violating the tetrahedron rule must remain 
unchanged, once the vacuum fluctuation has been created.

The energy associated with such a vacuum fluctuation is 
then 
\be
\Delta E_{\rm vac} = V + \epsilon({\bf k}) + \bar\epsilon(-{\bf
k})~~~~. 
\label{eq:DeltaEvac}
\en
the potential energy lowered by the kinetic energy of the two defects
(here we differentiate between the kinetic energy $\epsilon({\bf k})$
of the defect associated with the three d$^1$ sites, and the 
$\bar\epsilon({\bf k})$ of a square 
containing three d$^0$ sites).
If the net energy of {\it any} vacuum fluctuation is negative, the
tetrahedron rule will break down, and at least in the 
crudest approximation, a metal--insulator transition will 
take place.  Numerically, for spinless
fermions on a checkerboard lattice (of size eight to sixteen 
squares, with periodic boundary conditions), 
we find that this occurs for $V \approx 7t$.  


The structure of the vacuum fluctuation is such that a spin
$1/2$ chain with an {\it even} number of sites is linked symmetrically to a
Heisenberg ring (see Fig. \ref{fig4}), implying a singlet ground state. 
Importantly, the $+e/2$ ``tail'' of the vacuum fluctuation 
(the tetrahedron with only one d$^1$ site),  
can annihilate with one of the fractional charges $-e/2$ 
produced by the fragmentation of an added electron.
This makes it possible to locally reassemble the added electron 
anywhere in the lattice.
Furthermore, fluctuations can recombine in such a way that they
create the objects associated with an extra electron and one hole (the latter
being an open chain of spins).  Therefore we expect a balance between 
fractional charges $\pm e/2$ and full charges $\pm e$ when electrons (holes) 
are added to the system, or even in the metallic state.

While the dynamics of the defect is relatively straightforward 
in the case of spinless fermions, the introduction of spin greatly 
complicates matters. 
The kinetic energy term conserves the spin of the 
mobile electrons, 
so the motion of a defect is accompanied by drastic rearrangements 
of the spin in the Heisenberg chains connected by hopping events.
We therefore speculate that at energies smaller than
the typical scale $J$ of spin excitations, the effective 
hopping of fractional charges may be reduced relative to
the (coherent) motion of electrons 
in the channels provided by empty sites. 

\section{Conclusions}

In this paper we have demonstrated that excitations with 
fractional charge $\pm e/2$ can occur in a pyrochlore lattice 
with strong short--range Coulomb repulsions.
They appear in pairs with a delocalized spin $1/2$ (electron) 
or spin $0$ (vacuum fluctuation).
Fractionally charged excitations were first proposed in the 
context of heavily doped trans--polyacetylene by Wu and Schrieffer
\cite{WuSch}, where they are linked to a commensurate
superstructure of the polymer chain.  
They have also famously been invoked in a ground--breaking paper 
by Laughlin \cite{Laughlin}, in connection with the fractional 
quantum Hall effect.
However, polyacetylene is a one--dimensional 
system, and the quantum Hall effect is a
phenomenon occurring in two--dimensions and high magnetic field.
To the best of our knowledge the present case provides the 
first example of a microscopic model supporting fractional 
charge in three dimensions.

In polyacetylene and related one--dimensional model systems, 
the prerequisite for the fractionalisation 
of charge is the existence of degenerate vacua.
The crucial prerequisite found here for the
appearance of fractional charges is the highly degenerate 
ground state of the pyrochlore lattice in the absence of 
kinetic energy terms. 
Just like a moving soliton in polyacetylene,
which transforms one of the two ground--state configurations
into the other, a moving fractional charge in a pyrochlore lattice 
links different degenerate configurations of the system. 
Furthermore, because at finite $V$ there exist a density
of order $(t/V)^2$ of vacuum fluctuations which can recombine
with fractional charges, the model also supports
some admixture of excitations carrying the full
electronic charge $-e$.

The above considerations apply primarily to pyrochlore insulators, 
when doped with electrons or holes.  An example of such a
system is magnetite (Fe$_3$O$_4$), an insulating spinel oxide,
whose B sites form a pyrochlore lattice of $Fe^{2+}$ and $Fe^{2+}$
ions.  We note, however, that spinel oxides are prone to 
structural phase transitions which lift the huge degeneracy of the 
states obeying the tetrahedron rule. 
The notable exception to this pattern is LiV$_2$O$_4$. 
Longer--range Coulomb repulsions will also act to reduce the 
degeneracy.   All of these effect will make it harder for
fractional charges to propagate freely.


Obviously, many unsolved problems remain.
For example, the calculation of thermodynamic and transport 
properties of the model, or the effect of an applied magnetic 
field, have yet to be attempted.
It would also be worthwhile to search for other lattice geometries 
in which fractional charges could occur. 
The present communication can only point to these problems 
and does not pretend to provide an answer.  It should,
however, be considered as a first step in that direction.

\vspace*{0.25cm} 
\baselineskip=10pt{\small \noindent This work was supported under the 
visitors program of MPI--PKS (K.\ P.\ and N.\ S.) }

%
%

\end{document}